\documentclass[copyright,creativecommons]{eptcs}
\providecommand{\event}{WFLP 2016} 

\usepackage{amssymb}
\usepackage{stmaryrd}
\usepackage{amsmath}
\usepackage{proof}
\usepackage{url}
\usepackage{xspace}
\usepackage{graphicx}
\usepackage{latexsym}

\renewcommand{\tt}{\ttfamily}
\newcommand{\codefont}{\small\tt}
\newcommand{\code}[1]{\mbox{\codefont{#1}}}
\newcommand{\ccode}[1]{``\code{#1}''}
\usepackage{listings}
\lstnewenvironment{curry}{}{}
\newcommand{\listline}{\vrule width0pt depth1.5ex}


\newcommand{\us}{\char95} 

\begin{document}
\sloppy

\title{A Typeful Integration of SQL into Curry}

\author{Michael Hanus \qquad\qquad Julia Krone
\institute{Institut f\"ur Informatik, CAU Kiel, Germany}
\email{mh@informatik.uni-kiel.de}
}
\def\titlerunning{A Typeful Integration of SQL into Curry}
\def\authorrunning{M. Hanus \& J. Krone}

\maketitle

\begin{abstract}
We present an extension of the declarative programming language Curry
to support the access to data stored in relational databases
via SQL.
Since Curry is statically typed, our emphasis on this SQL integration
is on type safety. Our extension respects the type system
of Curry so that run-time errors due to ill-typed data
are avoided.
This is obtained by preprocessing SQL statements at compile time
and translating them into type-safe database access operations.
As a consequence, the type checker of the Curry system
can spot type errors in SQL statements at compile time.
To generate appropriately typed access operations,
the preprocessor uses an entity-relationship (ER) model
describing the structure of the relational data.
In addition to standard SQL, SQL statements
embedded in Curry can include program expressions
and also relationships specified in the ER model.
The latter feature is useful to avoid the error-prone use
of foreign keys.
As a result, our SQL integration supports a high-level and
type-safe access to databases in Curry programs.
\end{abstract}

\section{Introduction}
\label{sec:introduction}

Many real world applications require access to databases
in order to retrieve and store the data they process.
To support this requirement, programming languages
must provide mechanisms to access databases, like relational
databases which we consider in this paper.
Relational database systems use SQL to access and manipulate data.
Since SQL is typically not part of the syntax of a programming language,
there are two principle approaches:
\begin{enumerate}
\item Pass SQL queries and statements as strings to some
database connection.
\item Provide a library of database access and manipulation operations
which are used inside the program and automatically
translated to SQL.
\end{enumerate}
Both approaches have their pros and cons.
Although the first approach is quite popular, like JDBC in Java,
it is well known that passing strings is a source of security leaks
if their syntax is not carefully checked,
in particular, in web applications \cite{Huseby03}.
Since these SQL strings are constructed at run time,
SQL syntax errors can occur during execution time.
Moreover, the types of the data obtained from the database
depend on the database scheme.
As a consequence, data must be checked
at run time, e.g., via type casts, whether they fit to the expected types.
Altogether, the ``string-based approach'' does not lead
to a reliable programming style.

These drawbacks are the motivation for the second approach, i.e.,
to develop specific database libraries in programming languages.
If a programmer uses only operations of these libraries
to communicate with a database system, syntax problems due to
ill-formed SQL queries cannot occur.
If the host programming language is strongly typed,
one could also avoid the aforementioned run-time type errors.
For instance, Haskell/DB \cite{LeijenMeijer99} provides
a set of monadic combinators to construct queries
as Haskell program expressions.
In this way, queries become program entities that can be statically
checked and combined with other program expressions.
Due to the use of a fixed set of combinators,
the expressiveness is limited compared to SQL.
SML\# \cite{OhoriUeno11} avoids this drawback
by extending the syntax and type system of Standard ML
to represent SQL statements as typed expressions.
These are checked
at compile time (by translating them into expressions that
use definitions of an SQL-specific library)
and translated at run time to SQL.
Hence, type safety is combined with a quite flexible type system.

One disadvantage of these library-based query embeddings
is the gap to the concrete and well known SQL syntax.
Programmers familiar with SQL must learn to use
the specific library operations or syntactic extensions
to obtain the same results as with the direct use of SQL.
Furthermore, the orientation towards the type of the data stored in
the database instead of higher-level data models, like
entity-relationship (ER) models, could be error prone if foreign keys
occur in relations.

To avoid these drawbacks, we propose a new intermediate approach.
We implemented this approach in the functional logic language
Curry \cite{Hanus97POPL,Hanus16Curry}, but it can also
be transferred to other languages with higher-order
programming features and a strong type system.
The basic ideas of our proposal are:
\begin{enumerate}
\item 
We allow the embedding of actual SQL queries in Curry programs.
Since SQL is not conform with the standard syntax of Curry,
these queries are embedded as ``integrated code,''
a concept which is recently supported by some Curry systems.
For instance, the following piece of code is a valid program:
\begin{curry}
studAgeBetween :: Int -> Int -> IO (SQLResult [(String,Int)])
studAgeBetween min max =
  ``sql Select Name, Age
        From Student Where Age between {min} and {max}
        Order By Name Desc;''
\end{curry}
This operation returns the name and age of all students stored in the
database with an age between the bounds \code{min} and \code{max}
provided as parameters,
where the result list is ordered descending by the names.
\item
To check the correctness of SQL queries at compile time,
they are analyzed by a preprocessor.
This preprocessor replaces SQL queries by calls to operations
of a library to support a typed access to databases.
The preprocessor takes an ER model of the database into account
in order to generate calls to appropriately typed operations.
\item
The ER model is also used to avoid the explicit use of foreign keys
in SQL queries. For this purpose, we slightly extend the syntax of SQL
and allow conditions on relations in SQL queries.
\item
The actual access to the database is done during run time
by translating the library operations and combinators
into SQL.
\end{enumerate}
As a result, we obtain a framework which allows to
embed SQL query strings into the program code, but these
queries are checked at compile time to avoid
possible run-time errors.

In the next section, we sketch some basic features of Curry
and the already existent framework to integrate foreign code
in Curry programs.
Section~\ref{sec:cdbi} surveys the Curry libraries
implementing a type-safe database interface which are used
in our transformation process.
Since we use the entity-relationship model to specify
the logical structure of the database, we describe in
Sect.~\ref{sec:erd} this model and how it is used in our framework.
The design and implementation of our SQL parser and transformation
is described in Sect.~\ref{sec:sqlparser} before we discuss
related approaches in Sect.~\ref{sec:related} and conclude.

\section{Curry and Integrated Code}
\label{sec:db-programming}

We assume familiarity with functional logic programming
\cite{AntoyHanus10CACM,Hanus13}
and Curry \cite{Hanus97POPL,Hanus16Curry} so that
we sketch only some basic concepts.
Note that we only use the functional kernel for the
database integration (in contrast to \cite{Hanus04JFLP})
so that a detailed understanding of the logic and concurrent
features of Curry is not required.

Curry is a modern declarative multi-paradigm language
which amalgamates the most important features
of functional and logic languages in order to provide a variety
of programming concepts to the programmer.
For instance, demand-driven evaluation, higher-order functions,
and strong polymorphic typing from functional programming
is combined with logic programming features, like
computing with partial information (logic variables),
constraint solving, and non-deterministic search.
It has been shown that this combination leads to
better abstractions in application programs
such as implementing graphical user interfaces \cite{Hanus00PADL},
web frameworks \cite{HanusKoschnicke14TPLP},
or access and manipulation of persistent data \cite{Hanus04JFLP}.

The syntax of Curry is close to Haskell \cite{PeytonJones03Haskell}
but allows also \emph{free} (\emph{logic}) \emph{variables}
in program rules and a non-deterministic choice of expressions.
For instance, the following program defines the well-known list
concatenation and an operation \code{perm} which
non-deterministically returns a permutation of the input list
(note that the last rule contains
a functional pattern \cite{AntoyHanus05LOPSTR}
to select some element of the input list):
\begin{curry}
(++) :: [a] -> [a] -> [a]        perm :: [a] -> [a]
[]     ++ ys = ys                  perm []            = []
(x:xs) ++ ys = x : (xs ++ ys)      perm (xs++[x]++ys) = x : perm (xs++ys)
\end{curry}
Higher-order declarative programming languages
support the embedding of domain specific languages (DSLs)
by defining powerful combinators.
For instance, this has been done in functional logic languages
for parser combinators \cite{CaballeroLopez99},
business models \cite{MazanekHanus11},
graphical user interface \cite{Hanus00PADL}, or
typed web interfaces \cite{Hanus06PPDP}.
However, there are also domains where a specific notation for DSLs
exists so that the transformation into a standard functional
syntax is undesirable. 
Examples of this kind are regular expressions, XML documents,
or SQL queries.
To support the direct use of such notations in Curry programs,
we implemented a ``code integrator'' for Curry
(which has some similarities to quasiquoting in Haskell \cite{Mainland07}).

\emph{Integrated code} is a string in a source program
which follows its own syntax rules and is translated into
a standard Curry expression by a preprocessor, also called
\emph{code integrator}.
To mark a piece of integrated code in a Curry program,
it must be enclosed in at least two back ticks and ticks.
One can also use any larger number of back ticks and ticks
(for instance, if a smaller number of them might occur in the integrated code),
but the number of starting back ticks and ending ticks
must always be identical.
After the starting back ticks, an identifier specifies
the kind of integrated code so that, in principle,
any number of different kinds of codes can be integrated
into a Curry program.
For instance, the current code integrator supports
with the code identifier \code{regex}
the POSIX syntax for regular expressions.
Thus, the following expressions are valid in source programs:
\begin{curry}
  if s ``regex (a|(bc*))+'' || s ````regex (ab*)+'''' then $\ldots$
\end{curry}
The code integrator replaces this code by a standard Curry expression
which matches a string against the given regular expression.
This is obtained by exploiting
a match operation and data structures
to specify regular expressions defined in the standard
library \code{RegExp}.\footnote{%
\url{http://www.informatik.uni-kiel.de/~pakcs/lib/RegExp.html}}
For instance, the second regular expression above is replaced
by the Curry expression
\begin{curry}
`match` [Plus [Literal 'a', Star [Literal 'b']]]
\end{curry}
Note that the Curry expression replaced for the integrated
code is put into a single line in order to avoid potential problems
with Curry's layout rule and to keep the line numbering of the
original program, which is important for meaningful error messages
of the front end subsequently invoked.

The integrated code of regular expressions allows
to apply the usual notation for regular expressions inside
Curry programs without extending the syntax of the base language.
As a further example, we specify a predicate which tests
whether a string has the structure of Curry identifiers
as follows:
\begin{curry}
isID :: String -> Bool
isID s = s ``regex [a-zA-Z][a-zA-Z0-9_']*''
\end{curry}
Currently, the code integrator, distributed with the Curry systems
PAKCS \cite{Hanus16PAKCS} and KiCS2 \cite{BrasselHanusPeemoellerReck11},
supports syntactic embeddings for regular expressions,
format printing (like C's \code{printf}),
HTML and XML (with layout rules to avoid the explicit writing
of closing tags), and, quite recently, an SQL dialect
which is described in this paper.

In order to allow an easy integration of new languages into the
code integrator, it provides a generic interface.
Basically, a language-specific parser must be implemented
as a function from an input string (containing the piece of
integrated code) to a translated output string (actually,
it is a monadic operation to provide for error messages, warnings, etc).
If such a function is implemented as a parser for the
specific language and a translator that yields a Curry expression,
it can be connected to the actual code integrator by registering
this parser.
Thus, the infrastructure to embed SQL query strings in Curry programs
is available. However, in order to get a type-safe translation
of SQL queries into Curry expressions, some library support
is necessary. This is sketched in the following section.

\section{Curry Database Interface Libraries}
\label{sec:cdbi}

To abstract from the concrete database and provide
abstractions for type-safe access to database entities,
we developed a family of libraries for this purpose.
Since they are the basis of the type-safe embedding of SQL into Curry,
we sketch the basic structure of these libraries in this section.

The Curry Database Interface (CDBI) libraries
implement different abstraction layers for accessing relational
databases via SQL.
On the base level, the libraries
implement the raw connection to a relational database.
Currently, only connections to SQLite databases are supported,
but other types of databases could easily be added.
The library defines an abstract data type \code{Connection}
and various operations on it. For instance, the operation
\begin{curry}
connectSQLite :: String -> IO Connection
\end{curry}
returns an open connection to an SQLite database with a given name.
To ease the handling of individual database accesses, there are
types
\begin{curry}
type SQLResult a = Either DBError a$\listline$
type DBAction a  = Connection -> IO (SQLResult a)
\end{curry}
Hence, \code{DBAction} is the type of operations that communicate
with the database, which could return an error or some data
(modeled by the type \code{SQLResult}).
Moreover, there are also monadic operations to combine database actions.
Beyond basic operations to read, write, and close (\code{disconnect})
the connection,
there are further operations to execute SQL commands on the connection
(which return the answers of the database), and begin, commit, or rollback
transactions.
To execute a database action on an SQLite database with a given name,
we define the following operation:\footnote{This simple implementation
closes the connection after each action.
In the actual implementation,
the open connection is stored instead of closing it
and re-used for the execution of subsequent database actions.}
\begin{curry}
runWithDB :: String -> DBAction a -> IO (SQLResult a)
runWithDB dbname dbaction = do conn <- connectSQLite dbname
                               result <- dbaction conn
                               disconnect conn
                               return result  
\end{curry}
The raw database access works with strings, e.g., numbers
stored in database entities are communicated in their string representation
between the database and the Curry program.
In order to set up a typeful communication, where values of particular
types are transmitted to the database and values of some expected
types are returned,
data types for the various kinds of SQL values and types are introduced:
\begin{curry}
data SQLValue = SQLString String | SQLInt  Int  | SQLFloat Float
              | SQLChar   Char   | SQLBool Bool | SQLDate ClockTime | SQLNull$\listline$
data SQLType = SQLTypeString | SQLTypeInt  | SQLTypeFloat
             | SQLTypeChar   | SQLTypeBool | SQLTypeDate
\end{curry}
Exploiting these types, one can define an operation to execute
a ``typed'' SQL query with the following type:
\begin{curry}
select :: String -> [SQLValue] -> [SQLType] -> DBAction [[SQLValue]]
\end{curry}
The first argument is the string of the SQL query which contains
some ``holes'' for concrete values (marked by \code{'?'}), like
\begin{curry}
"select * Student where Age = '?' and Name = '?';"
\end{curry}
The second argument contains SQL values for these holes
and the third argument are the types expected as return values
so that the entire call is a database action which returns a table,
i.e., a list of rows of the expected values.
The usage of holes in the SQL string instead of concrete values
is useful to ensure safe encoding of all SQL values and, thus,
avoid security leaks known as SQL injections.
As an example, the following expression describes a database action
to extract the age and email address of a student:
\begin{curry}
select "select Age,Email from Student where First = '?' and Name = '?';"
       [SQLString "Joe", SQLString "Fisher"]
       [SQLTypeInt, SQLTypeString]
\end{curry}
Although this typed select statement is a good starting point,
its use requires the correct application of the column types
of the database.
Since each database stores entities of a particular kind
(according to its database scheme) and these entities
can be related to types of the Curry program,
the CDBI libraries define a type to specify this
relationship:
\begin{curry}
data EntityDescription a = ED String [SQLType]
                              (a -> [SQLValue]) ([SQLValue] -> a)
\end{curry}
Hence, an entity description consists of the name of the entity,
a list of column types, and conversion functions from a Curry type
to SQL values and back.
For instance, if we define a type to represent a student
(consisting of the name, first name, matriculation number, email, and age)
by
\begin{curry}
data Student = Student String String Int String Int
\end{curry}
then a corresponding entity description could be defined as follows:
\begin{curry}
studentDescription :: EntityDescription Student
studentDescription =
  ED "Student"
     [SQLTypeString, SQLTypeString, SQLTypeInt, SQLTypeString, SQLTypeInt]
     (\ (Student name first matNum email age)
       -> [SQLString name, SQLString first, SQLInt matNum,
            SQLString email, SQLInt age])
     (\ [SQLString name, SQLString first, SQLInt matNum, SQLString email,
         SQLInt age] -> Student name first matNum email age)
\end{curry}
Note that this must not be done by the programmer
since the data type and entity description can be automatically
derived from the ER model of the database (see next section).

Beyond the typed access to database entities, it is also
important to ensure type-safe queries containing
selection criteria (i.e., the \ccode{where} part of SQL select statements).
Thus, there is a data type to describe selection criteria
(with an optional group-by clause):
\begin{curry}
data Criteria = Criteria Constraint (Maybe GroupBy)
\end{curry}
The \code{Constraint} data type allows to define a logical combination of
comparisons of values from database columns or constants.
The type to describe such values is:
\begin{curry}
data Value a = Val SQLValue | Col (Column a)
\end{curry}
Hence, a value used in a criteria is either a constant
or a database column of a particular type.
To ensure a type-safe use of such values, the constructor \code{Val}
is not directly used but there is a family of constructor functions, like
\begin{curry}
int :: Int -> Value Int
int = Val . SQLInt
\end{curry}
The columns are introduced with their entity descriptions, e.g.,
the age column of a student entity has type
(where the concrete definition contains the column and entity names):
\begin{curry}
studentColumnAge :: Column Int
\end{curry}
The data type \code{Constraint} contains a number of constructors
to describe the possible conditions in \code{where} clauses.
Moreover, there are constructor functions for various constraints,
like the greater-than relation:
\begin{curry}
(.>.) :: Value a -> Value a -> Constraint
\end{curry}
Hence, we can express the criteria that the age of a selected student
should be greater than 21 as:
\begin{curry}
Col studentColumnAge .>. int 21
\end{curry}
The complete list of all possibilities to construct criteria
is defined in the module \code{CDBI.Criteria}.\footnote{%
\url{http://www.informatik.uni-kiel.de/~pakcs/lib/Database.CDBI.Criteria.html}}
Thanks to these abstractions,
a constructed selection criteria is ensured to be type safe.
For instance, it is compile-time type error
when a database column of type \code{Int} is compared with a string.

Based on the \code{select} operation described above,
there is a higher-level operation \code{getEntries} to
access entities with some given criteria and ordering options.
For instance, the following expression
selects the first five students who are older than 21 ordered by their names
in descending alphabetical order (if the first argument is \code{Distinct}
instead of \code{All}, then only distinct results are returned):
\begin{curry}
getEntries All
           studentDescription
           (Criteria (Col studentColumnAge .>. int 21) Nothing)
           [descOrder studentColumnName]
           (Just 5)
\end{curry}
In order to implement queries that return joins of several entities,
the CDBI libraries allow to combine entity descriptions and
provide operations to retrieve such combined entities from a
database. For the selection of single columns of one or more tables,
the libraries provide functions similar to \code{getEntries}.
Furthermore, there are specific operations for
database insertions, updates and deletion of entries.

The CDBI libraries are the base to implement type-safe SQL queries.
Each SQL statement in a Curry program will be translated into
appropriate calls to the CDBI operations sketched above.
However, this requires a precise description of
the types of the database scheme.
In order to abstract from foreign keys (to be more precise, we
want to support a type-safe handling of foreign keys),
we use a description of the database scheme in the form
of an entity-relationship model, as discussed in the following section.

\section{Entity-Relationship Models}
\label{sec:erd}

\begin{figure*}[t]
\begin{center}
  \includegraphics[trim= 0 320 120 0,scale=0.7, clip=true]{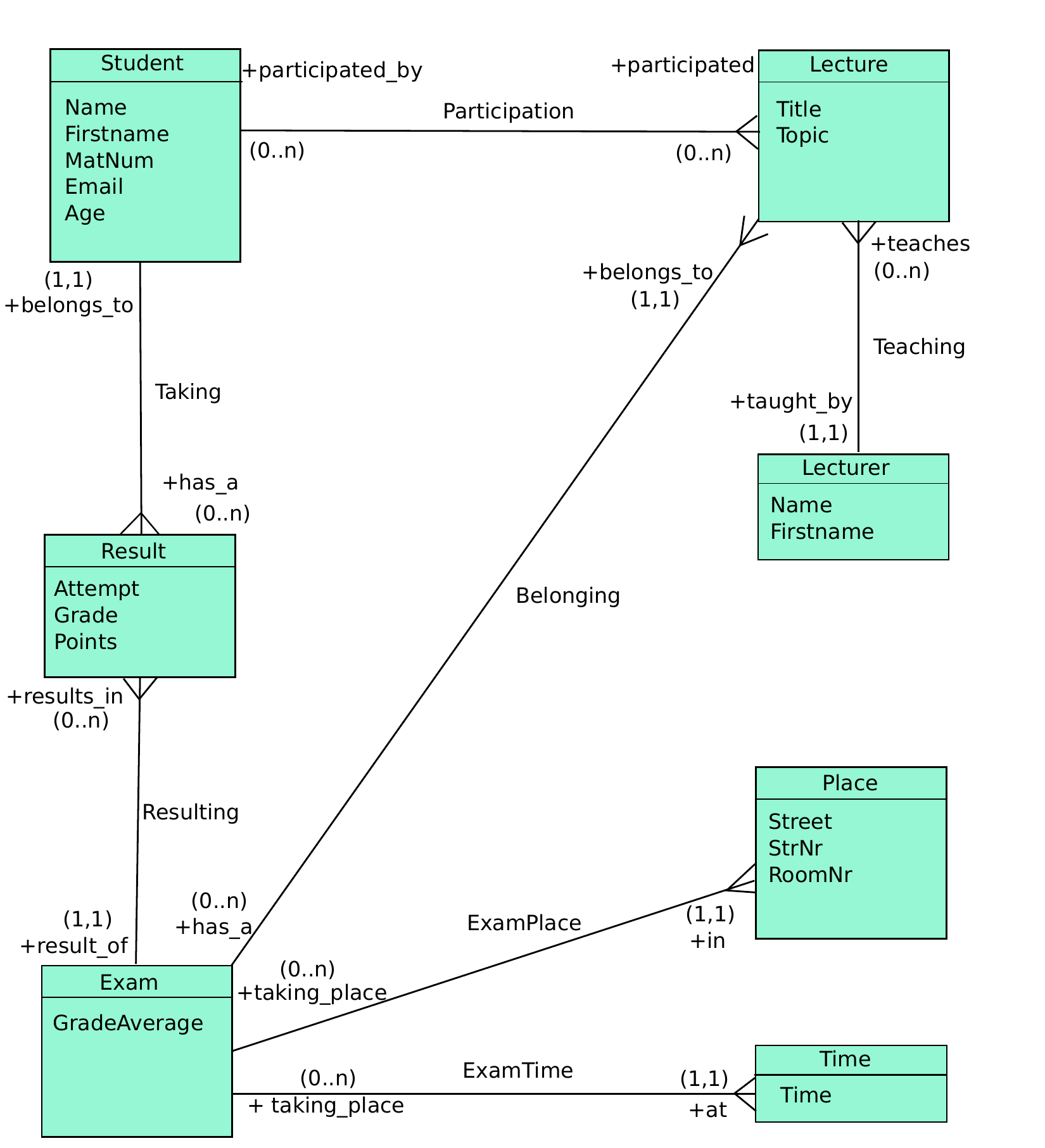}
\end{center}
\caption{A simple entity-relationship diagram for university lectures}
\label{fig:erd}
\end{figure*}

The entity-relationship model \cite{Chen76} describes
the structure and specific constraints of data stored in a
relational database.
It uses a graphical notation, called entity-relationship diagrams (ERDs),
to visualize the conceptual model.
The data of an application is modeled by entities that have attributes
and relationships between entities.
Relationships have cardinality constraints
that must be satisfied in each valid state of the database, e.g.,
after each transaction.
Figure~\ref{fig:erd} shows a concrete ERD which we use in our examples.

In order to be independent of a concrete ER modeling tool,
we use a textual representation of ERDs as Curry data terms
defined in \cite{BrasselHanusMueller08PADL}.
In this representation, an ERD consists of a name (also used as the module name
in the generated code) and lists of entities and relationships:
\begin{curry}
data ERD = ERD String [Entity] [Relationship]
\end{curry}
Each entity consists of a name and a list of attributes, where
each attribute has a name, a domain, and specifications
about its key and null value property:
\begin{curry}
data Entity    = Entity String [Attribute]
data Attribute = Attribute String Domain Key Null
\end{curry}
Similarly, each relationship consists of a name and the names
and cardinalities of the related entities
(more details can be found in \cite{BrasselHanusMueller08PADL}).

Since the SQL queries embedded in a Curry module are related
to a particular ER model (which can be specified in a compiler directive
in the head of a module), the given ER model is preprocessed
to make it available to the final Curry program.
This is the task of a specific translation tool (\code{ERD2CDBI})
which performs the following three steps:
\begin{enumerate}
\item 
The ER model is transformed into tables of a relational database,
i.e., the relations of the ER model are either represented
by adding foreign keys to entities (in case of (0/1:1) or (0/1:n) relations)
or by new entities with the corresponding relations
(in case of complex (n:m) relations).
This task is performed by the existing \code{erd2curry} tool
and described in detail in \cite{BrasselHanusMueller08PADL}.
After this transformation, a relational database with the resulting structure
and constraints is created.
\item
A Curry module is generated containing the definitions of
entities and relationships as data types.
Since entities are uniquely identified via a database key,
each entity definition has, in addition to its attributes, this key as the
first argument.
For instance, the following definitions are generated
for our university ERD (among many others):
\begin{curry}
data StudentID = StudentID Int$\listline$
data Student = Student StudentID String String Int String Int$\listline$
-- Representation of n:m relationship Participation:
data Participation = Participation StudentID LectureID
\end{curry}
Note that the two typed foreign key columns (\code{StudentID}, \code{LectureID})
ensures a type-safe handling of foreign-key constraints.

Moreover, for each entity, an entity description and definitions
for their columns are generated as described in Sect.~\ref{sec:cdbi}.
\item
Finally, an \emph{info file} containing information about all entities,
attributes and their types, and relationships is created.
This file provides the SQL parser and translator (see next section) with
the necessary information to generate appropriate CDBI library calls.
The info file contains a data term of the following type:
\begin{curry}
data ParserInfo = PInfo String String RelationTypes NullableFlags
                        AttributeLists AttributeTypes 
\end{curry}
The first and second components specify the absolute
path to the database (later used to for the database connection)
and the name of the Curry module containing
the data types and operations generated from the ER model as sketched above.
The latter is necessary to generate correctly qualified function calls
and types in the translated SQL code.

The third component (\code{RelationTypes}) contains information
about all relations of the ER model, like names, related entities,
and cardinality. This information is used by the SQL translator
to translate constraints on relationships into database constraints relating
foreign keys.

The fourth component (\code{NullableFlags}) assigns to each column name
a Boolean flag which is true if this column can contain null values.
Note that a column of type $\tau$ with possible null values is represented
in Curry by the type \code{Maybe$\;\tau$}. In order to generate
correct types for accessing column values from the Curry program,
the information about null values is required by the SQL translator.

The fifth component (\code{AttributeLists})
maps each table name to a list of their column names.
The final component (\code{AttributeTypes}) maps each column name
to its type, an information which is obviously required for our
type-safe translation of SQL queries.

For the sake of a simple textual representation of this information,
the components of this SQL parser information are stored as lists
in the generated info file. When the parser reads this file,
the information is converted into finite maps
to support a more efficient access inside the SQL parser.
\end{enumerate}

\section{SQL Parsing and Translation}
\label{sec:sqlparser}

\subsection{Design Goals}

Now we have all components to implement the integration of
SQL queries inside a Curry program.
As mentioned in the introduction,
the following piece of code should be valid in a Curry program:
\begin{curry}
studGoodGrades :: IO (SQLResult [(String, Float])
studGoodGrades = ``sql Select Distinct s.Name, r.Grade 
                       From   Student as s, Result as r
                       Where  Satisfies s has_a r And r.Grade < 2.0;''
\end{curry}
This operation connects to the database and returns,
if no error occurs, a list of pairs containing the
names and grades of students having a grade better than \code{2.0}.
Note that this query is beyond pure SQL since it also includes
a condition on the relation \code{has\us{}a} specified in the ER model
(\ccode{Satisfies s has\us{}a r}).

It should also be possible to include Curry expressions instead
of constants in conditions of SQL queries.
In order to distinguish them from SQL code,
embedded Curry expressions are enclosed in curly braces.
For instance, the following operation returns the email addresses
of all students with a given last name (or terminates with an error):
\begin{curry}
studEmail :: String -> IO [String]
studEmail name = do dbresult <- ``sql Select s.Email From Student as s
                                      Where s.Name = {name};''
                    return (either (error . show) id dbresult)
\end{curry}
These examples demonstrate the basic objectives of our
approach that must be implemented by the SQL parser and
translator:
\begin{enumerate}
\item 
For easy usability, standard SQL notation is allowed
(instead of specific database access operations as shown
in Sect.~\ref{sec:cdbi}).
\item
SQL statements are type safe, i.e., every type inconsistency
between the Curry program and the database tables is detected
at compile time. For instance, if we declare the type of
\code{studEmail} as
\begin{curry}
studEmail :: Int -> IO [String]
\end{curry}
a type error will be reported by the Curry compiler.
\item
In addition to standard SQL conditions, we also support
a condition \ccode{Satisfies} to check relationships between entities
w.r.t.\ the ER model.
\item
The SQL embedding is deep, i.e., we can write embedded SQL queries
in any place of a Curry program (of course, only if the type
is appropriate in the context of the SQL query)
and we can also embed Curry expressions of
appropriate types in SQL queries.
\item
From the embedded SQL query, a standard SQL query is generated
and transmitted
at run time so that it will be processed by the database system.
\end{enumerate}

\subsection{Structure of the Implementation}

\begin{figure*}[t]
\begin{center}
  \includegraphics[trim= 0 210 0 0, scale=0.5, clip=false]{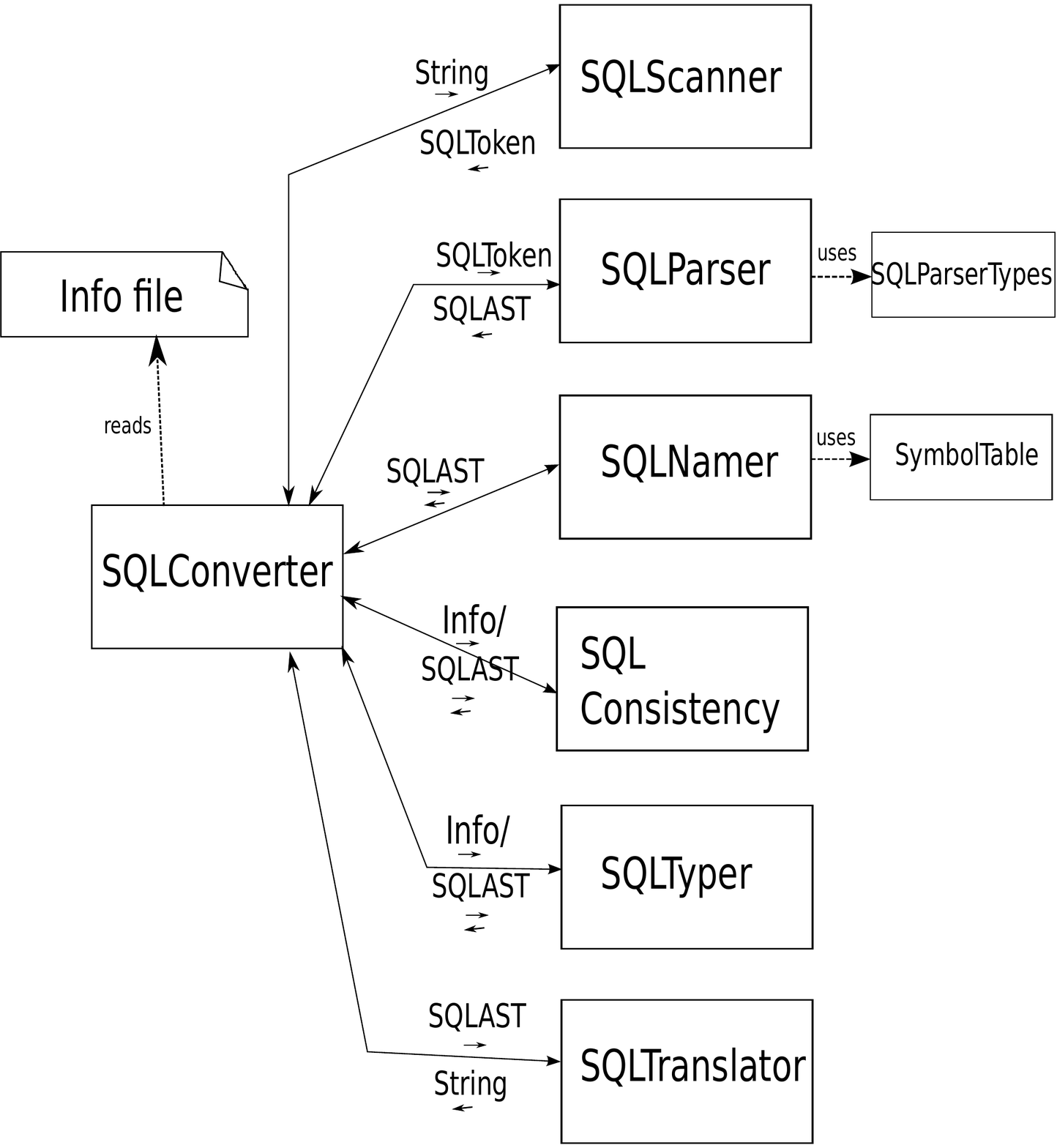}
\end{center}
\caption{Structure of the SQL compiler}
\label{fig:sqlcompiler}
\end{figure*}

The basic structure of the implemented SQL compiler is depicted
in Fig.~\ref{fig:sqlcompiler}.
As shown in this figure, the compiler uses the info file,
described in Sect.~\ref{sec:erd}, to generate appropriate
Curry expressions for the SQL strings.
Basically, our SQL compiler follows the typical structure of compilers.
For instance, in the first phase, an abstract syntax tree (AST)
of the SQL statement is generated.
The subsequent phases correspond to the semantic analysis
and code generation and are sketched in the following.

The \code{SQLNamer} is the first stage of the semantic analysis.
It uses a specific symbol table to resolve and replace
pseudonyms for data tables (introduced by the keyword \ccode{as}
in an SQL statement) by the full table names.
Furthermore, table references are numbered in order
to generate correct calls to CDBI operations (e.g., if the
same table is referenced several times in an SQL statement).
Errors are thrown if table pseudonyms can not be resolved.
This happens in case a pseudonym was defined but not used,
a pseudonym was defined for more than one table,
or if the pseudonym was not defined or is not visible.

The \code{SQLConsistency} module implements
various consistency checks.
For instance, all table, column, and relationship names occurring
in the SQL statement must be part of the ER model.
Other consistency checks are related to the use of relations
in the specific \code{Satisfies} condition and the correct usage of null values
in insert statements.
Furthermore, it is checked that null values are not used in conditions
except in the predefined functions \code{Is Null} and \code{Not Null},
since this is generally forbidden by the SQL standard.

The \code{SQLTyper} implements an initial type checking phase.
For all expressions in an SQL statement where
at least one column reference is involved as well as for constant values,
the types are available and, thus, can be checked.
Therefore, possible type errors might be reported already by
the SQL preprocessor and not just during by the subsequent
compilation of the Curry compiler.
For instance, the SQL query
\begin{curry}
``sql Select s.Name From Student as s Where s.Age = 20.5;''
\end{curry}
will not be passed to the Curry compiler, because the
preprocessor reports the error message
\begin{curry}
$\ldots$ Type error: Int (Age) and Float are not compatible.
\end{curry}
However, if a condition contains an embedded Curry expression instead of
a constant value, the preprocessor can not ensure type safety
at this stage, since
the types of Curry expressions are not available during preprocessing.
Instead, the \code{SQLTyper} deduces the type required
for the Curry expression from its context in the query.
This information is used in the final phase of our compiler
to generate a type-correct call of the corresponding
CDBI library function.
Hence, the strong typing of the CDBI library, as discussed in
Sect.~\ref{sec:cdbi}, ensures the type consistency of SQL statements
even in the presence of embedded Curry expressions.
For instance, consider the definition
\begin{curry}
studNamesWithAge x = ``sql Select s.Name
                           From   Student as s
                           Where  s.Age = {x};''
\end{curry}
Although the preprocessor has no information about the type of \code{x},
which will be eventually inferred by the subsequent invocation
of the Curry compiler, it can deduce that the embedded Curry expression
\code{\{x\}} must have type \code{Int} since it is compared
with the \code{Int}-valued column \code{Age}.
Thus, the preprocessor translates the \code{Where} condition into
the following CDBI call:
\begin{curry}
equal (Col studentColumnAge) (int x)
\end{curry}
Since the CDBI function \code{int} requires an argument of
type \code{Int} (see Sect.~\ref{sec:cdbi}),
the Curry compiler would report a type error if
\code{x} has not type \code{Int}. Hence, the Curry compiler
infers the type
\begin{curry}
studNamesWithAge :: Int -> IO (SQLResult [String])
\end{curry}
and reports an error message if \code{studNamesWithAge} is
declared with the following type:
\begin{curry}
studNamesWithAge :: String -> IO (SQLResult [String])
\end{curry}
Currently, it is not allowed to compare two embedded expressions, as in
\begin{curry}
  ``sql Select $\ldots$ Where {x*2} = {y};''
\end{curry}
Since the preprocessor is not able to infer the required types of
these embedded expressions, appropriate CDBI calls cannot be
generated.
This is not a principle restriction of our approach but
due to the current type system of Curry.
If the base language supports overloading (like Haskell or
an extension of Curry with type classes \cite{Martin-Martin11}),
the preprocessor can generate a call to an overloaded
CDBI operation.
Another option is a complete re-implementation of the
type inference of the base language in the SQL preprocessor
(as done in \cite{OhoriUeno11}).
Since we are interested in a modular SQL integration that could
be used with different implementations of Curry,
we do not follow the latter option.
This seems a reasonable compromise
since such kinds of SQL conditions are usually not necessary.

The final stage of the SQL preprocessor is the translation of the
SQL AST into a valid Curry expression by the \code{SQLTranslator} module.
Since the SQL AST is decorated with all necessary type information
by the previous phases, the code generation is somehow
straightforward. However, the translation of conditions
on relationships (\code{Satisfies}) is technically involved
since one has to consider the various kinds of relationships
and how they are represented with foreign keys in the relational
database.

To show a concrete translation example, consider the
query operation \code{studNamesWithAge} defined above.
Our SQL compiler generates the following Curry code
(which is reformatted for readability):
\begin{curry}
studNamesWithAge x = runWithDB "/$\ldots$/Uni.db"
   (getColumn []
      [SingleCS All
         (singleCol studentNameColDesc 0 none)
         (TC studentTable 0 Nothing)
         (Criteria (equal (colNum studentColumnAge 0) (int x)) Nothing)]
      [] Nothing)
\end{curry}
Although there are various further parameters in order to cover
the different features of SQL, the basic structure is visible.
Since the CDBI libraries eventually translate
the Curry calls to a single SQL query, the features of the
database systems are exploited to answer the query.
For instance, the call \ccode{studNamesWithAge 30}
results in the following SQL query:
\begin{curry}
select ("Student"."Name") from 'Student'
                          where (("Student"."Age") == 30);
\end{curry}
The complexity of the translation process becomes more visible
when several relations and foreign keys are used.
For instance, our initial query in this section is translated
into the following Curry code:
\begin{curry}
studGoodGrades :: IO (SQLResult [(String, Float)])
studGoodGrades = runWithDB "Uni.db"
   (getColumnTuple []
      [TupleCS Distinct
         (tupleCol (singleCol studentNameColDesc 0 none)
                   (singleCol resultGradeColDesc 0 none))
         (TC studentTable 0 (Just (crossJoin, TC resultTable 0 Nothing)))
         (Criteria (And [equal (colNum studentColumnKey 0)
                               (colNum resultColumnStudentTakingKey 0),
                         lessThan (colNum resultColumnGrade 0) (float 2.0)])
                   Nothing)]
      [] Nothing)
\end{curry}
This complex expression is translated into the compact SQL query:
\begin{curry}
select Distinct ("Student"."Name"), ("Result"."Grade")
from  'Student' cross join 'Result'
where (("Student"."Key") == "Result"."StudentTakingKey")
      and (("Result"."Grade") < 2.0)  ;
\end{curry}
It should be obvious that a distinct advantage of our
SQL embedding is the handling of foreign keys,
which is type-safe (actually, hidden in the SQL query)
in our framework but error prone with raw SQL.

Although we discussed only the translation of \code{select} queries,
our SQL embedding also supports other SQL statements, like
\code{insert}, \code{delete}, or \code{update} statements.
In contrast to \code{select} queries, which have the result
type \code{IO$\;$(SQLResult$\;$[$\tau$])} (where $\tau$ depends
on the kind of selected data), the latter SQL statements
are of type \code{IO$\;$(SQLResult$\;$())}.
Due to the complexity of SQL, our compiler does not support
the full SQL syntax. In particular, the use of join operators
(e.g., natural joins) is limited.
A specification of the supported SQL syntax as an EBNF grammar and
more details about the implementation of the SQL compiler
can be found in \cite{Krone15}.

\section{Related Work}
\label{sec:related}

A popular approach to integrate SQL into
a programming language is the string-based approach,
like JDBC in Java. The programmer writes SQL statements as
string constants in the program (or implement a method to construct
them during run time) which are passed to some database connection.
As already discussed,
this does not ensure type safety and might cause security leaks.
Therefore, we consider in the following only approaches
that try to avoid these problem by using specific abstractions
of the base programming language, in particular,
of declarative programming languages.

Since relations stored in a relational database
can be considered as facts defining a predicate of a logic program,
logic programming provides a natural framework
for connecting databases (e.g., see \cite{Das92,GallaireMinker78}).
Unfortunately, the well-developed theory in this area
is not accompanied by practical implementations.
For instance, Prolog systems rarely come
with a standard interface to relational databases.
An exception is Ciao Prolog which has a persistence module \cite{CorreasEtAl04}
that allows the declaration of predicates as facts that
are persistently stored in a relational database.
This hides the communication with the database and
supports a simple method to query the relational database
by logic programming methods.
However, due to the untyped nature of Prolog,
type safety is not ensured and queries are not explicitly
distinguished from updates that are handled by predicates with side effects.
A similar concept, but with a typed data access and
a clear separation between queries and updates, has been proposed in
\cite{Hanus04JFLP} for Curry.

The latter concept has been extended in \cite{BrasselHanusMueller08PADL},
where an entity-relationship (ER) model of the database
is used to provide safe update operations that respect
the constraints of the ER model and avoid the explicit use
of foreign keys.
Although this ensures type safety similar to our approach,
the strongly typed combinators do not support all SQL constructs,
in particular, no complex joins.
Hence, the programmer must implement more complex queries
using features of the programming language.
While this is easily possible with higher-order operations
like \code{map} and \code{filter},
this approach does not exploit the optimization features
of database systems.

A disadvantage of such library-based approaches,
like \cite{LeijenMeijer99} for Haskell
or \cite{BrasselHanusMueller08PADL} for Curry,
is the fact that programmers must know, in addition to the base
language and SQL, the specific methods and operations
to implement SQL statements with these libraries.
This can be avoided by a deeper embedding of SQL into the base
language.
A recent approach \cite{OhoriUeno11,OhoriEtAl14}
does this for Standard ML (SML).
The language SML\# extends to syntax of SML to
write SQL queries as program expressions.
In order to provide meaningful types to these expressions,
the type system is extended with record polymorphism
so that a principal type can be inferred for a type-correct
SQL expression. This type is also used to check
the type consistency of a concrete database with a given SQL query.
The extension of the syntax and the type system of the base language
requires the (re-)implementation of an SML compiler,
in contrast to our lightweight preprocessor approach
which can be combined with any Curry compiler.
Due to the representation of SQL queries as SML\# expressions,
there are syntactic differences between SQL queries
and their SML\# equivalents.
In our approach, these differences are avoided by considering
SQL statements as integrated code instead of true Curry code.
The use of record polymorphism in SML\# supports the static typing
of SQL expressions without a given database scheme.
On the other hand, we use a high-level ER model of the database
to type foreign keys (on the CDBI library level) and
avoid the use of foreign keys in SQL queries.
Our small syntactic extension of SQL avoids potential errors
in the use of foreign keys and increases the expressiveness
of queries similarly to the Ruby on Rails framework.

A strongly typed database access is also an objective
of other functional languages integrating database query facilities,
like Kleisli \cite{Wong00} or Links \cite{CooperLindleyWadlerYallop06}.
In contrast to our approach, these languages do not offer
SQL syntax to formulate queries but require the use of other
language constructs (e.g., list comprehensions)
to formulate queries that are transformed into SQL.

\section{Conclusions}
\label{sec:conclusions}

We presented a framework to support a high-level and reliable
access to relational databases from programs written
in the declarative programming language Curry.
The characteristic features of our framework are:
\begin{itemize}
\item 
Database access and manipulation of persistent data is formulated in
a slight extension of SQL. Hence, it is easy to use
since most programmers are already familiar with SQL.
Moreover, one can exploit high-level features of SQL,
like complex join operations or constraints.
\item
In order to detect ill-formed or ill-typed SQL statements
at compile time, the SQL statements are preprocessed
and translated into type-safe operations of a specific
database access library.
\item
The library operations generate the corresponding SQL code
at run time. Hence, the format of embedded data can be checked
so that typical security leaks, like SQL injections, are avoided.
\item
Since the preprocessor uses a logical model of the database
specified as an ER model, the use of relationships
of this model inside conditions of SQL queries is supported.
Hence, the potentially error-prone use of (untyped) foreign keys is avoided.
\end{itemize}
To implement this framework, we exploited some available
infrastructure, in particular, to inject foreign code into
Curry programs.
The implementation of this preprocessor
is available in the current releases of the Curry systems
PAKCS \cite{Hanus16PAKCS} and KiCS2 \cite{BrasselHanusPeemoellerReck11}.
In principle, such a framework could also be
implemented for other strongly typed programming languages.
Nevertheless, the declarative and high-level nature of Curry
eased our implementation of this non-trivial compilation task.

For future work, it might be useful to support further database systems.
In order to increase run-time reliability,
it would be reasonable to check the
consistency of the ER model used in the SQL compiler
with the actual database scheme, e.g., when compiling
a program, or before every program execution.

\paragraph{Acknowledgements.}
The authors are grateful to Jasper Paul Sikorra for
the implementation of the foreign code integrator
and Mike Tallarek for implementing the initial version of
the CDBI libraries.

\bibliographystyle{eptcs}
\bibliography{mh}

\end{document}